\documentclass[letterpaper, 10 pt, conference]{ieeeconf}
\IEEEoverridecommandlockouts                              
\usepackage{cite}
\usepackage{array,pdfsync,enumerate,tikz,enumerate}
\usepackage[dvips]{epsfig}
\usepackage{amssymb,amsmath}
\usepackage{caption}
\usepackage{subcaption}
\usepackage{multirow}
\usepackage{color}
\definecolor{linkcol}{rgb}{0,0.0,1.0}
\definecolor{citecol}{rgb}{0.0,0.6,0.0}

\IEEEoverridecommandlockouts
\usepackage{cite}
\usepackage{amsmath,amssymb,amsfonts}
\usepackage{algorithmic}
\usepackage{graphicx}
\usepackage{textcomp}
\usepackage{hyperref}
\hypersetup{colorlinks,breaklinks,
	linkcolor=linkcol,urlcolor=citecol,
	anchorcolor=linkcol,citecolor=citecol,bookmarks=true}
\usepackage{caption}
\usepackage{subcaption}
\newtheorem{theorem}{Theorem}[section]

\newtheorem{proposition}[theorem]{Proposition}
\newtheorem{remark}[theorem]{Remark}
\newcommand{\rpos}{\mathbb{R}_{\geq 0}}

\newcommand{\tx}[1]{\text{#1}}

\usepackage{graphicx}      % include this line if your document contains figures
\usepackage{cite}
\begin{document}
	
	\title{Generation of Wheel Lockup Attacks on Nonlinear Dynamics of Vehicle Traction}
	
	\author{Alireza Mohammadi, Hafiz Malik, and Masoud Abbaszadeh% <-this % stops a space
		%\thanks{This work was not supported by any organization}% <-this % stops a space
		\thanks{This work is supported by NSF Award 2035770. A. Mohammadi and H. Malik are with the
			Department of Electrical and Computer Engineering, University of Michigan-Dearborn, MI 48127 USA. M. Abbaszadeh is with 
			GE Global Research, NY 12309 USA.  
			Emails: {\tt\small \{amohmmad,hmalik\}@umich.edu, abbaszadeh@ge.com}. Corresponding Author: A. Mohammadi.}%
	}

\maketitle		
		
		\begin{abstract}                % Abstract of not more than 250 words.
			 There is ample evidence in the automotive cybersecurity literature that the car brake ECUs can be maliciously reprogrammed. 
			 Motivated by such threat, this paper investigates the capabilities of an adversary who can directly control the frictional brake actuators and would like to induce wheel lockup conditions leading to catastrophic road injuries. This paper demonstrates that the adversary despite having a limited knowledge of the tire-road interaction characteristics has the capability of driving the states of the vehicle traction dynamics 
			 to a vicinity of the lockup manifold in a finite time by means of a properly designed attack policy for the frictional brakes. This 
			 attack policy relies on employing a predefined-time controller and a nonlinear disturbance observer acting on the wheel  slip error dynamics. Simulations under various road conditions demonstrate the effectiveness of the proposed attack policy.          
		\end{abstract}
		
	%	\begin{keyword}
	%	quadratic programming, autonomous robots, nonlinear control systems, saturation, guidance systems.
	%	\end{keyword}

%	\end{frontmatter}

%~~~~~~~~~~~~~~~~~~~~~~~~~~~~~~~~~~~~~~~~~
%\include{PhDBiblio}	
%===============================================================================
%
% ShimResearch folder:
%
% Teixeira: 
%One of the attack scenarios analyzed corresponds to a novel
%type of detectable attack, the bias injection attack. Although this
%attack may be detected, it can drive the system to an unsafe region
%and it only requires limited model knowledge and no information
%about the system state.
%~~~~~~~~~~~~~~~~~~~~~~~~~~~~~~~~~~~~~~~~~	
%
    \section{Introduction}
%
% Documentation of attacks and motivating the ECU takeover: 
Electronic Control Units (ECUs), which are an integral part of the modern automotive control systems,  are embedded devices that are typically networked together on one or more buses and communicate with each other based on protocols such as CAN and FlexRay~\cite{huang2018vehicle}.  The automotive control systems including their in-vehicle networks (IVNs) and ECUs can be attacked in different ways~\cite{kim2021cybersecurity,avatef2017}. Indeed, there is a plethora of well-documented attacks to access the IVNs (see, e.g., the work by Koscher, Chekoway and collaborators~\cite{checkoway2011comprehensive,koscher2020experimental}).   One type of attack, which can be carried out using interfaces such as the vehicle OBD port, Wi-Fi, and cellular networks, can be used to reprogram the vehicle ECUs and hence manipulating the car physical behaviors including its steering and braking. For instance, Miller and Valasek~\cite{miller2015remote} carried out such an attack, which was effective on all Fiat-Chrysler vehicles, by reprogramming a V850 chip to disable the car braking system (also, see~\cite{froschle2017analyzing} for other attacks on the brake ECUs).  % Having access to the IVN, the attacker can easily sniff the CAN packets, which %are broadcast to all components on the CAN bus, and/or inject CAN packets onto the CAN bus while masquerading as a legitimate car ECU. 

Motivated by the possibility of malicious reprogramming of the car braking ECUs, this paper investigates the capabilities of an adversary who has taken over the braking ECUs and would like to induce wheel lockup conditions during braking. Wheel lockup can cause   severe degradation in vehicle steerability, directional stability, and general control over
the car~\cite{gillespie1992fundamentals} leading to catastrophic road injuries~\cite{giummarra2021classification}. 

There exists a rich body of literature on designing robust cyber-physical attacks against  linear time-invariant (LTI) networked control systems including zero-dynamics attacks on non-minimum phase systems (see, e.g.,~\cite{teixeira2015secure,hoehn2016detection,park2016adversary,park2019stealthy}) and bias injection attacks on power plants and smart grids (see, e.g.,~\cite{kontouras2017impact,wang2020detection}). The intent of the adversary is to drive the system to an unsafe operating region while requiring limited model knowledge of the system dynamics in both types. 

\begin{figure}
	\centering
	\includegraphics[scale=0.38]{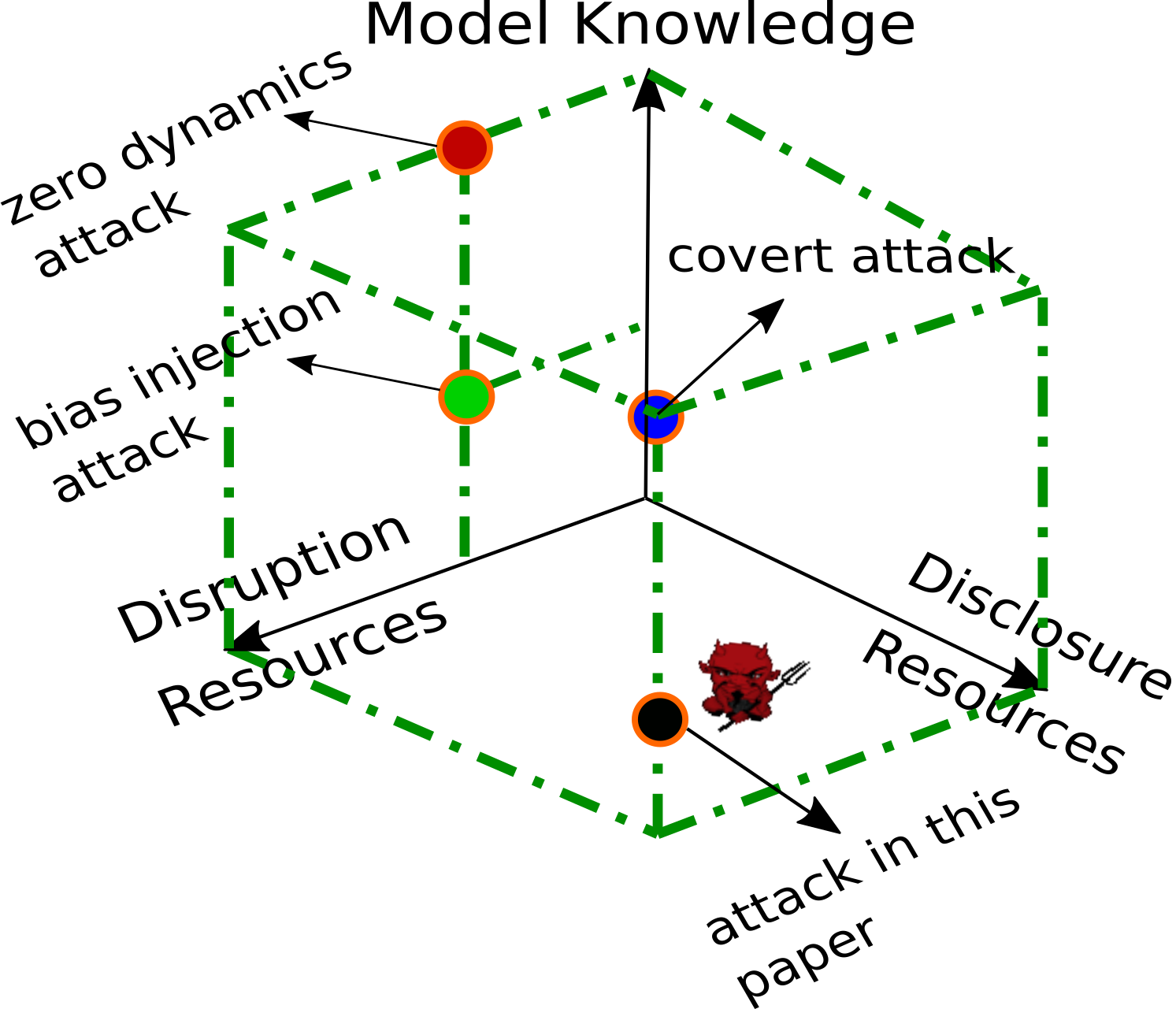} 
	\vspace{-1ex}
	\caption{Classification of the proposed attack scheme within the cyber-physical attack space. The adversary has complete authority over the 
	brake actuators (full disruption resources) and can read the states of the vehicle traction dynamics (full disclosure resources) from the in-vehicle network, e.g., the CAN bus. However, the adversary does have a very limited knowledge of the plant dynamics and its interactions with the ambient environment (vehicle traction dynamics and its interaction with the road).}
	\label{fig:classifyAttack}
	\vspace{-1cm}
\end{figure}

In this paper, we demonstrate that the adversary having taken over the brake ECUs and possessing a limited knowledge of the tire-road interaction characteristics can induce  lockup on the vehicle wheels by means of a properly designed attack policy. The  attack policy relies on employing a predefined-time controller~\cite{sanchez2015predefined} for driving the vehicle states to a vicinity of the lockup manifold and a nonlinear disturbance observer (NDOB) that will compensate for the limited knowledge of the adversary (see, e.g., the references~\cite{li2014disturbance,chen2004disturbance,mohammadi2017nonlinear} and applications of NDOBs in bipedal robots~\cite{mohammadi2018hybrid} and power plants~\cite{rigatos2021flatness}). According to 
Teixeira \emph{et al.}'s widely-accepted taxonomy~\cite{teixeira2015secure} we are concerned with \emph{physical aspects} of such an attack. In other words, this paper considers direct perturbation of the vehicle traction dynamics with malicious intents under the assumption that the adversary has direct control over the brake actuators. Therefore, the proposed attack policy can be classified within the attack space due to Teixeira \emph{et al.}~\cite{teixeira2015secure} as in Figure~\ref{fig:classifyAttack}.   
% summary of the contributions  

\noindent  \textbf{Contributions of the Paper.} As the first contribution, this paper adds to the body of knowledge on physical attack generation against nonlinear dynamical systems where the vast literature on attack generation  mostly considers cyber-physical systems with linear dynamics. One of the very few exceptions is the work by Park \emph{et al.}~\cite{park2018stealthiness} where a stealthy zero-dynamics attack against a quadruple-tank process is designed. Second, this paper provides an affirmative answer to the question of whether an adversary with limited knowledge of the vehicle traction dynamics and the tire-road interaction  characteristics can induce wheel lockup conditions in a finite time interval. Such an answer provides important insights for the emerging area of attack generation against platoons of vehicles (see, e.g.,~\cite{merco2020hybrid}). Finally, this paper contributes to  
improving the performance of pre-defined time controllers for single-input single-output systems in the presence of uncertainties and disturbances by means of adding an NDOB-based feedforward compensation term to the control action. 

The rest of this paper is organized as follows. After presenting some preliminaries in Section~\ref{sec:prelims}, we present the nonlinear dynamical model of the vehicle traction in Section~\ref{sec:dynmodel}. Next, we formulate the wheel lockup attack policy objective under uncertain tire-road friction characteristics in Section~\ref{sec:AttackPolicy}. Thereafter,  in Section~\ref{sec:attackDesign}, we present  our attack policy based on using predefined-time controllers and nonlinear disturbance observers. After presenting the simulation results in Section~\ref{sec:sims} under various road conditions, we conclude the paper with final remarks and future research directions in Section~\ref{sec:conc}.
    \section{Preliminaries}
\label{sec:prelims}

In this section we provide some preliminary definitions from finite-time stability theory. The interested reader is referred to~\cite{polyakov2014stability,zhou2016asymptotic,sanchez2015predefined,polyakov2015finite,song2019time} for further details. 

Consider the dynamical system
\begin{equation}
\dot{y} = f(t,y;\zeta)
\label{eq:prelim1}
\end{equation}
where $t\in [t_0, \infty)$ for some $t_0>0$, $y\in \mathbb{R}^n$, and $\zeta \in \mathbb{R}^m$ denote time, the state, and the parameters of the system in~\eqref{eq:prelim1}, respectively. We denote the solutions to~\eqref{eq:prelim1} starting from $y_0=y(t_0)$  by $y(t,y_0)$. We say that the non-empty set $\Gamma \subset \mathbb{R}^n$ is \emph{locally finite-time stable}  for~\eqref{eq:prelim1} if $\Gamma$ is locally asymptotically stable for~\eqref{eq:prelim1} and there exists a neighborhood of $\Gamma$, denoted by $\mathcal{N}(\Gamma)$, such that any solution $y(t,y_0)$ of~\eqref{eq:prelim1} with $y_0\in \mathcal{N}(\Gamma)$  reaches $\Gamma$ in a finite time moment $t^\ast=T(y_0)$, where the \emph{settling-time function} $T: \mathcal{N}(\Gamma) \to \rpos$ is locally bounded. We say that~\eqref{eq:prelim1} is \emph{locally fixed-time stable} if it is locally finite-time stable and there exists some positive constant $T_{max}$, called the \emph{settling-time}, such that $T(y_0) \leq T_{max}$ for all $y_0\in \mathcal{N}(\Gamma)$. If $\mathcal{N}(\Gamma) = \mathbb{R}^n$, the definitions will become global. 

%We denote the set of non-negative real numbers by $\mathbb{R}_{\geq 0}$. Following the notation in~\cite{zhou2016asymptotic}, given the interval $J=[t_0,\infty)\subset \mathbb{R}_{\geq 0}$ where $t_0$ is some positive real number, we let $\mathbb{PC}(J,\mathbb{R})$ denote the union of the sets of all real-valued piecewise continuous and continuously differentiable functions defined on $J$. Consider the LTV system 
%%
%\begin{equation}
%\dot{x} = a(t) x(t), 
%\label{eq:prelim2}
%\end{equation}
%% 
%where $a(.) \in \mathbb{PC}(J,\mathbb{R})$ and $x\in \mathbb{R}$. We say that the LTV system in~\eqref{eq:prelim2} is \emph{uniformly exponentially stable} if there exist positive constants $\gamma$ and $k$ such that $|x(t)| \leq k |x(t_0)| \exp(-\gamma (t-t_0) )$ for all $t, t_0 \in J$ with $t\geq t_0$.
 
%
    \section{Vehicle Traction Dynamical Model}
\label{sec:dynmodel} 
In this section, we provide a brief overview of the single-wheel model of rubber-tired vehicles under straight-ahead braking conditions. The presented dynamic model can capture the steady and transient tractive performance while demonstrating how a vehicle can undergo stable braking or lockup~\cite{olson2003nonlinear,johansen2003gain,de2012torque,li2018hierarchical}. Conventionally, the tire/wheel rate of  rotation and the forward vehicle speed are taken as dynamic states. Accordingly, the quarter-car dynamics governing the vehicle longitudinal motion during braking are given by (see, e.g.,~\cite{johansen2003gain,de2012torque})
\begin{subequations}\label{eq:tractionDyn1}
	\begin{align}
	\dot{v} = -g_\alpha \mu(\lambda) - \frac{\Delta_v(t,v)}{M},\\
	\dot{\omega} = \frac{M g_\alpha r}{J} \mu(\lambda) - \frac{T_a}{J} - \frac{\Delta_w(t,\omega)}{J}, 
	\end{align}
\end{subequations}
\noindent where $M$, $r$, and $J$  denote the quarter-car mass, wheel radius, and wheel inertia, respectively.  Furthermore, the vehicle speed $v$ and the wheel rotational speed $\omega$ during braking belong to $\mathcal{D}_b:=\{ (v,\omega) | v > 0, \; 0 \leq r\omega \leq v \}$. The braking torque $T_a$ is the input to the dynamical system in~\eqref{eq:tractionDyn1}. Moreover, the longitudinal slip $\lambda$ is given by 
\begin{equation}\label{eq:slipDefn}
\lambda := \frac{v- r\omega}{\max(v, r\omega)}.
\end{equation}

\begin{figure}
	\centering
	\includegraphics[scale=0.65]{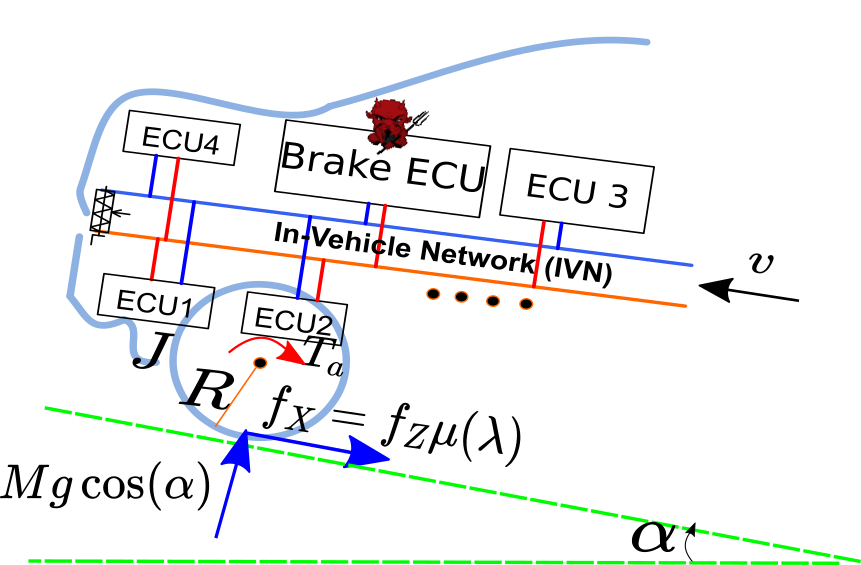} 
	\caption{Schematic of the single-wheel braking model.}
	\label{fig:dynModel}
\end{figure}
During braking, when $(v,\omega) \in \mathcal{D}_b$, we have $\lambda = \tfrac{v- r\omega}{v}$. Therefore, $\lambda \in [0,1]$. We let the constant $g_\alpha$ denote $g \cos(\alpha)$ where $\alpha$ is the road slope. Finally, $\mu(\lambda)$, $\Delta_v(t,v)$, and $\Delta_w(t,\omega)$ denote the uncertain nonlinear friction coefficient, the force, and the torque disturbances resulting from unmodeled dynamics, respectively. 

\begin{remark}
	\label{rem:mu}
	There are a variety of ways to represent the function $\mu(\cdot)$ including the Magic Formula and Burckhardt representation (see, e.g.,~\cite{de2011optimal}). Equations like Magic Formula are empirical equations based on fitting coefficients that are widely utilized for modeling the interaction between the tire tread and road pavement, where the longitudinal force arising from this interaction is given by the Magic Formula. In the propositions that follow in Section~\ref{sec:attackDesign}, we do not assume any particular closed-form representation for the nonlinear friction coefficient function $\mu(\cdot)$ and only assume that $\mu : \Lambda \to \mathbb{R}$ is a continuous function on the closed interval $\Lambda := [0,1]$. Accordingly,  $\mu(\cdot)$ attains its maximum $\mu_{\tx{max}}$ and minimum $\mu_{\tx{min}}$ on the closed interval $\Lambda$ because of the well-known properties of continuous functions on compact sets.
\end{remark}

As is customary in the vehicle dynamics literature (see, e.g.,~\cite{de2012torque}), it will be assumed that the disturbances $\Delta_v(t,v)$, and $\Delta_w(t,\omega)$ satisfy the following uniform bounds
\begin{eqnarray}\label{eq:distBound}
\nonumber
|\Delta_v(t,v)| \leq \bar{\Delta}_v,\;  |\Delta_w(t,\omega)| \leq \bar{\Delta}_\omega, \\
\text{ for all } (t,v,\omega) \in [0,\infty) \times \mathcal{D}_b. 
\end{eqnarray}

Under the change of coordinates $(v,\omega) \mapsto (v, \lambda)$, the longitudinal dynamics become (see, e.g.,~\cite{de2012torque,olson2003nonlinear} for the 
details of derivation) 
\begin{subequations}\label{eq:tractionDyn2}
	\begin{align}
	\dot{v} &= -g_\alpha \mu(\lambda) - \frac{\Delta_v(t,v)}{M},\\
	\dot{\lambda} &= \frac{g_\alpha}{v}\big\{ (\lambda - 1 - \nu) \mu(\lambda) + \Upsilon_a + \Upsilon_{\Delta, w} + (\lambda -1)\Upsilon_{\Delta, v} \big\}, 
	\end{align}
\end{subequations}
\noindent where $\nu:=\tfrac{M R^2}{J}$ is the dimensionless ratio of vehicle to wheel inertia, $\Upsilon_a := \frac{r}{J g_\alpha} T_a$ is the dimensionless brake torque, and $\Upsilon_{\Delta, w} := \frac{r}{J g_\alpha} \Delta_w(t, \omega)$, $\Upsilon_{\Delta, v} := \frac{\Delta_v(t,v)}{M g_\alpha}$ are the dimensionless force and torque disturbances acting on speed and slip dynamics, respectively. As it can be seen from~\eqref{eq:tractionDyn1} and~\eqref{eq:tractionDyn2}, the wheel slip $\lambda$ couples 
the tire/wheel dynamics with that of the vehicle speed. In $(v, \lambda)$ coordinates, we have  
\begin{equation}
\mathcal{D}_b=\big\{ (v,\lambda) | v > 0, \; \lambda \in \Lambda := [0,1] \big\}.
\label{eq:Db}
\end{equation}

 Using the brake input $\Upsilon_a$, the adversary would like to induce unstable braking conditions corresponding to lockup. The most severe case of lockup happens when $\lambda=1$. Therefore, following Olson \emph{et al.}~\cite{olson2003nonlinear}, we define the lockup manifold in the following way
\begin{equation}
\mathcal{W}_b^L := \big\{ (v,\lambda) \big| v>0,\, \lambda = 1 \big\}.
\label{eq:lockupManifold}
\end{equation}  
    \section{Attack Policy Objectives}
\label{sec:AttackPolicy}
In this section we formulate the attack policy objectives and state our assumptions about the adversarial knowledge of 
the vehicle dynamical model, the adversarial disruption resources, and the adversarial disclosure resources. 

\noindent\textbf{The Adversarial Knowledge.} Following  the notation by Teixeira \emph{et al.}~\cite{teixeira2015secure}, we denote the vehicle traction dynamics in~\eqref{eq:tractionDyn2} by $\mathcal{P}$. Moreover, we let the attacker's \emph{a priori}  knowledge model $\hat{\mathcal{P}}$ be given by 
\begin{subequations}\label{eq:tractionDyn2approx}
	\begin{align}
	\dot{v} = -{g}_{\alpha} \hat{\mu}(\lambda),\\
	\dot{\lambda} = \frac{{g}_\alpha}{v}\big\{ (\lambda - 1 - \hat{\nu}) \hat{\mu}(\lambda) + \hat{\Upsilon}_a \big\}, 
	\end{align}
\end{subequations}
\noindent where the adversary has no \emph{a priori knowledge} of the dimensionless torque and force disturbances $\Upsilon_{\Delta, w}$, $\Upsilon_{\Delta, v}$. Furthermore, the adversary has only an approximate knowledge of the tire-road interaction characteristics given by $\hat{\mu}(\lambda)$. When the adversary does not have any knowledge of $\mu(\lambda)$, the friction coefficient $\hat{\mu}(\lambda)$ is set equal to $0$ in $\hat{\mathcal{P}}$.  

\noindent\textbf{The Adversarial Disruption Resources.} To model the adversarial disruption resources, we assume that the reference malicious command generated by the attack policy $\hat{\Upsilon}_a$ passes through the following first-order delay system (see, e.g.,~\cite{de2012torque,li2018hierarchical})  
\begin{equation}
\tau_f \dot{\Upsilon}_a  = - \Upsilon_a + \hat{\Upsilon}_a(t- \delta_f), 
\label{eq:torqueResp}
\end{equation}
to generate the frictional braking torque response $\Upsilon_a$, which then gets applied to the traction dynamics in~\eqref{eq:tractionDyn2}.  It is remarked that the attacker does not have any knowledge of either the friction brake time constant $\tau_f$ or the friction brake deadtime $\delta_f$. In designing our attack policies in the next section, we assume that $\tau_f \approx 0$ and $\delta_f \approx 0$. However, the simulation results in Section~\ref{sec:sims} demonstrate the effectiveness of the attack policies when these assumptions do not hold. 

\noindent\textbf{The Adversarial Disclosure Resources.} We assume that the adversary knows and/or can compute the vehicle velocity as well as the wheel slip. This scenario corresponds to having complete access to disclosure resources in the cyber-attack space~\cite{teixeira2015secure} (also, see  Figure~\ref{fig:classifyAttack}). Indeed, as demonstrated in an experimental wireless attack against Tesla vehicles~\cite{nie2017free}, an adversary who has the capability of reprogramming the firmware of ECUs through the Unified Diagnostic Services (UDS) codified in ISO-14229~\cite{iso14229} can also read live data from the IVN (such as vehicle speed or engine speed). 

\noindent\textbf{Wheel Lockup Attack Policy Objective.} Given the vehicle longitudinal dynamics $\mathcal{P}$ in~\eqref{eq:tractionDyn2}, the attacker's \emph{a priori} knowledge $\hat{\mathcal{P}}$ in~\eqref{eq:tractionDyn2approx}, and the  braking response in~\eqref{eq:torqueResp}, design an attack policy $\hat{\Upsilon}_a$ such that the trajectories of the vehicle longitudinal dynamics during braking converge to any sufficiently close neighborhood of the lockup manifold  $\mathcal{W}_b^L$ within a finite time interval. 
     \section{Attack Policy Design}
\label{sec:attackDesign}
In this section, we present an attack policy in \emph{two steps} that can achieve the wheel lockup objective formulated in the previous section. We start with a simple attack policy based on the concept of predefined-time controllers proposed in~\cite{sanchez2015predefined} and assume that the adversary has an almost perfect knowledge of the vehicle dynamics. Then, we start adding to the sophistication level of the attack policy analysis and design by assuming less knowledge of the dynamical model. 

\noindent\textbf{Step 1: Design of predefined-time controllers.} Following the notation in~\cite{sanchez2015predefined}, we let 
\begin{equation}
\Phi_p(x):=\frac{\exp(|x|^p)}{p} |x|^{1-p} \tx{sign}(x),  
\label{eq:phip}
\end{equation}
for any $x\in \mathbb{R}$ and some real constant $0 < p < 1$, and   
\begin{equation}
\Phi_1(x):=\exp(|x|) \tx{sign}(x),
\label{eq:phi1}
\end{equation}
for any $x\in \mathbb{R}$. Furthermore, we define the lockup error as
\begin{equation}
\label{eq:eL}
e_L := \lambda - 1. 
\end{equation}
\noindent Therefore, if $e_L=0$ and $v>0$, the wheel is locked and $(v,\lambda) \in \mathcal{W}_b^L$. As it will be shown in this section, if the attack objective is met, the wheel will be locked in finite time. Hence, the vehicle speed will satisfy  
\begin{equation}
\label{eq:speedBound}
v \in [v_\tx{min}, v_\tx{max}], 
\end{equation}
during a successful attack, for some positive $v_\tx{min}$ and $v_\tx{max}$. 
%Using the functions defined in~\eqref{eq:phip} and~\eqref{eq:phi1}, we define the following 
%feedback control laws 
%%
%\begin{eqnarray}
%u_{np}^a(e_\tx{L}) &:=& -k_a \tx{sign}(e_L) -\frac{1}{T_c} \Phi_p(e_L), 
%\label{eq:unp}
%\\
%u_{n1}^a(e_\tx{L}) &:=& -\big(\frac{1}{T_c} + k_a \big) \Phi_1(e_L),
%\label{eq:un1}
%\end{eqnarray}
%%
%\noindent where $k_a$, $T_c$, and $0<p<1$  are some positive real constants. 
%
\begin{proposition}
	\label{prop:prop1}
	Consider the vehicle longitudinal dynamics $\mathcal{P}$ in~\eqref{eq:tractionDyn2} with the attacker's \emph{a priori} knowledge $\hat{\mathcal{P}}$ in~\eqref{eq:tractionDyn2approx} and the frictional braking response given by~\eqref{eq:torqueResp} with $\tau_f \approx 0$ and $\delta_f \approx 0$.  Suppose $\Upsilon_{\Delta, w}=0$, $\hat{\nu}=\nu$, and $\hat{\mu}(\cdot)=\mu(\cdot)$. Additionally, assume that the uniform bound on $\Delta_v(t,v)$ given by~\eqref{eq:distBound} holds. Given any positive constant $T_c$, the attack policy
	\begin{equation}
	\label{eq:attPolicyNoDob}
	\hat{\Upsilon}_a = \frac{v}{g_\alpha} u_{np}^a(e_\tx{L}) + \hat{\nu} \hat{\mu}(\lambda),  
	\end{equation}
	with $u_{np}^a(e_\tx{L})=  -(\tfrac{1}{T_c} + k p ) \Phi_p(e_L)$, where $0 < p < 1$, $k \geq k^{\ast\prime}:=\tfrac{M g_\alpha \mu_{\max} + \bar{\Delta}_v}{M v_{\tx{min}}}$, and $\Phi_p(\cdot)$ given by~\eqref{eq:phip}, makes the lockup manifold $\mathcal{W}_b^L$  globally finite-time stable with settling-time $T_c$. 
\end{proposition}

\begin{proof}
	The closed-loop dynamics of $e_L$ under the stated assumptions become  
\begin{equation}\label{eq:errorDyn0}
\dot{e}_L  =   u_{np}^a(e_\tx{L})  + \Delta_e(t,e_L),  
\end{equation}
\noindent where $\Delta_e(t,e_L) = \tfrac{g_\alpha}{v} e_L \{\mu(\lambda) + \Upsilon_{\Delta, v} \}$ is a vanishing 
perturbation satisfying $|\Delta_e(t,e_L)| \leq \tfrac{M g_\alpha \mu_{\max} + \bar{\Delta}_v}{M v_{\tx{min}}} |e_L|$. Therefore, 
by Lemma~4.1 in~\cite{sanchez2015predefined}, the proposition is proved. 
\end{proof}

Proposition~\ref{prop:prop1} assumes an almost perfect knowledge of the vehicle's model, where the only unknown is the 
disturbance force $\Delta_v(t,v)$ acting on the vehicle speed dynamics in~\eqref{eq:tractionDyn2}. The next proposition 
removes these restrictions further.    

\begin{proposition}
	\label{prop:prop2}
	Consider the stated assumptions in Proposition~\ref{prop:prop1} with $\hat{\nu}$ arbitrary, and     $\Delta_w(t,\omega)$, $\Delta_v(t,v)$ satisfying~\eqref{eq:distBound}. Furthermore, assume  that $\hat{\mu}: \Lambda \to \mathbb{R}$ is a continuous function. Then, given $0 < T_c < 1$, the attack policy~\eqref{eq:attPolicyNoDob} with $u_{np}^a(e_\tx{L})=  -k_a \tx{sign}(e_L) -\tfrac{1}{T_c} \Phi_p(e_L)$, $0 < p < 1$, $\Phi_p(\cdot)$ given by~\eqref{eq:phip}, and $k_a \geq k^\ast$ in which  
	\begin{equation}
	\label{eq:kBound}
		k^\ast := \frac{M g_\alpha \mu_{\max} + \bar{\Delta}_v}{M v_{\tx{min}}} + \frac{g_\alpha}{v_{\tx{min}}}\big( \hat\nu \hat\mu_{\text{max}} + \nu \mu_{\text{max}} + \frac{r \bar{\Delta}_w}{J g_\alpha} \big), 
	\end{equation}
	makes the lockup manifold $\mathcal{W}_b^L$ globally finite-time stable with settling-time $T_c$. 
\end{proposition}

\begin{proof}
	The closed-loop dynamics of $e_L$ under the stated assumptions read as 
	\begin{equation}\label{eq:errorDyn2}
	\dot{e}_L  =   u_{np}^a(e_\tx{L})   + \Delta_e^{\prime}(t,e_L),   
	\end{equation}
	\noindent where 
	\begin{eqnarray}
	\nonumber
	\Delta_e^{\prime}(t,e_L) & = & \frac{g_\alpha}{v} e_L \big\{\mu(\lambda) +  \Upsilon_{\Delta, v}  \big\} + \cdots \\
    &  & \frac{g_\alpha}{v} \big\{\hat{\nu}\hat{\mu}(\lambda) - \nu\mu(\lambda)\big\} + \frac{g_\alpha}{v} \Upsilon_{\Delta, w},
	\label{eq:kbound2}
	\end{eqnarray}
	is a non-vanishing perturbation. Due to the definition of wheel slip, we have $|e_L| \leq 1$ for all values of $\lambda \in \Lambda$. Therefore, it follows that $|\Delta_e^{\prime}(t,e_L)| \leq k^\ast$, where $k^\ast$ is given by~\eqref{eq:errorDyn2}. Consequently, 
	by Lemma~4.3 in~\cite{sanchez2015predefined}, the proposition is proved. 
\end{proof}

Similar to the line of argument in Proposition~\ref{prop:prop2} and using Lemma 4.2 in~\cite{sanchez2015predefined}, the following proposition, whose proof is omitted for the sake of brevity, removes the restrictions on the settling-time in Proposition~\ref{prop:prop2}. 

\begin{proposition}
Consider the stated assumptions in Proposition~\ref{prop:prop2}. Given any positive constant $T_c$, the attack policy 
\begin{equation}
\label{eq:attPolicyNoDob3}
\hat{\Upsilon}_a = \frac{v}{g_\alpha} u_{n1}^a(e_\tx{L}) + \hat{\nu} \hat{\mu}(\lambda),  
\end{equation}
with $u_{n1}^a(e_\tx{L})=  -(\tfrac{1}{T_c} + k_a) \Phi_1(e_L)$, where $k_a \geq k^\ast$, $k^\ast$ given by~\eqref{eq:kBound}, and $\Phi_1(\cdot)$ given by~\eqref{eq:phi1}, makes the lockup manifold $\mathcal{W}_b^L$ globally finite-time stable with  settling-time $T_c > 0$.  
\label{prop:prop3}	
\end{proposition}
%
%
%\begin{eqnarray}\label{eq:errorDyn}
%\nonumber
%\dot{e}_L & =   \frac{g_\alpha}{\hat{g}_\alpha} u_n^a(e_\tx{L}) - \frac{g_\alpha}{v} \hat{d}_a  + \frac{g_\alpha}{v} e_L \big\{\mu(\lambda) + \Upsilon_{\Delta, v}  \big\} + \\
%& \frac{g_\alpha}{v} \big\{\hat{\nu}\hat{\mu}(\lambda) - \nu\mu(\lambda)\big\} + \frac{g_\alpha}{v} \Upsilon_{\Delta, w} 
%\end{eqnarray}
%locally finite-time convergent to a compact set $\Omega_b^L$ containing $\mathcal{W}_b^L$ such that $|e_L| \leq f(| d |_\infty)$. 
%
\noindent\textbf{Step 2: Design of NDOBs for wheel slip error dynamics.} Thus far, the presented family of attack policies in Propositions~\ref{prop:prop1}--\ref{prop:prop3} depend on some \emph{a priori} knowledge of the vehicle  longitudinal dynamics $\mathcal{P}$ in~\eqref{eq:tractionDyn2}. In order to remove the need for such knowledge, we make our attack input dynamic by means of 
adding a feedforward compensation term to the proposed attack policies. In particular, we extend the brake attack policies in~\eqref{eq:attPolicyNoDob} and~\eqref{eq:attPolicyNoDob3} according to 
\begin{equation}
\label{eq:attPolicyDob}
\hat{\Upsilon}_a = \frac{v}{g_\alpha} u_{ni}^a(e_\tx{L}) + \hat{\nu} \hat{\mu}(\lambda) - \hat{d}_a, \; i=1,\, p, 
\end{equation}
where $\hat{d}_a \in \mathbb{R}$ is the output of the following NDOB (see, e.g.,~\cite{chen2004disturbance,mohammadi2017nonlinear} for details of derivation)
\begin{subequations}\label{eq:dobDyn}
	\begin{align}
	\dot{z}_a &= -L_d z_a - L_d \big\{ u_{ni}^a + \frac{g_\alpha}{v} (-\hat{d}_a + 
	\hat{\nu}\hat{\mu}(\lambda)  + p_a) \big\},\\
	\hat{d}_a &= z_a + p_a, 
	\end{align}
\end{subequations}
\noindent where $z_a \in \mathbb{R}$ is the state of the NDOB, and the relationship $p_a = L_d e_L$ between $L_d$, namely, the 
NDOB gain, and $p_a$, namely, the NDOB auxiliary variable, holds. 
%
%\begin{equation}
%\label{eq:pLrel}
%p_a = L_d e_L,   
%\end{equation}
%
\noindent Therefore, it follows that $L_d = \tfrac{\partial p_a}{\partial e_L}$. 

\begin{figure*}[t]
	\label{tab:simTab}
	\hspace{-1ex}
	\begin{subfigure}{0.45\textwidth}
		\centering
		\includegraphics[scale=0.40]{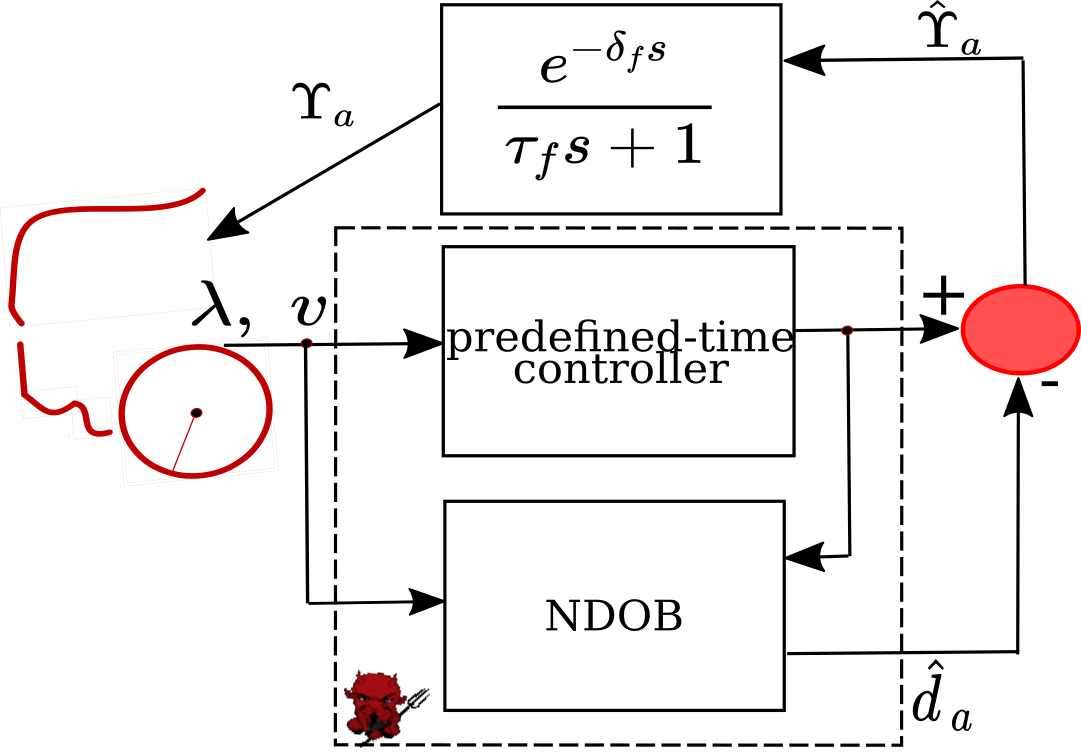} 
		\caption{}
		\label{fig:blockDiag}
	\end{subfigure}
%\hspace{2cm}
\begin{subfigure}{0.45\textwidth}
	{\scriptsize
		\begin{tabular}{ |c|c|c|c| } 
			\hline
			Quarter-car parameters & Road parameters & Attack policy parameters \\
			\hline
			$M=250$ kg & $\alpha=0^{\deg}$ & $T_c=0.95$ \\ 
			$\tau_f = 16$ ms & $c_1 = 1.28$, $c_1^\prime = 0.86$ & $p=0.15$ \\ 
			$\delta_f = 15$ ms & $c_2 = 23.99$, $c_2^\prime = 33.82$ & $k = 0$ \\ 
			$J = 1.5$ kg.m\textsuperscript{2} & $c_3 = 0.52$, $c_3^\prime = 0.35$ & $L_d = 2.65$ \\
			$R = 0.3$ m & -- & $\hat{\nu} = \nu = 15$ \\
			-- & -- & $\hat{\mu}(\lambda)=0$, $\tau_f=0$, $\delta_f=0$ \\
			\hline
		\end{tabular}
	}
	\caption{}
	\label{tab:simTab}
\end{subfigure}
\caption{(a) The proposed attack policy block diagram; (b)  numerical simulation parameters.}
\end{figure*}

The NDOB in~\eqref{eq:dobDyn}  has been designed based on considering the wheel slip error dynamics 
\begin{equation}\label{eq:errorDyn3}
\dot{e}_L  =   u_{ni}^a(e_\tx{L})   + \Delta_e^{\prime}(t,e_L) - \hat{d}_a, \, i=1,\, p,   
\end{equation} 
with the lumped disturbance $\Delta_e^{\prime}(t,e_L)$ given by~\eqref{eq:kbound2}. The NDOB output $\hat{d}_a$, which is being calculated by~\eqref{eq:dobDyn}, tries to cancel the lumped disturbance $\Delta_e^{\prime}(t,e_L)$. In the ideal case, when $\hat{d}_a \approx \Delta_e^{\prime}(t,e_L)$, complete disturbance cancellation is achieved and it appears to the predefined-time controller, which was designed in Step~1, that it is controlling a system with no disturbances. The block diagram of the proposed attack policy is depicted in Figure~\ref{fig:blockDiag}. 

The convergence properties of the disturbance tracking error $\hat{d}_a - \Delta_e^{\prime}$ are well-studied in the literature (see, e.g.,~\cite{li2014disturbance,chen2004disturbance,mohammadi2017nonlinear}) and for the sake of brevity we refer the readers to the aforementioned references. 

\begin{remark}\label{rem:choice}
	The NDOB in~\eqref{eq:dobDyn} has only one dynamic state and it does not rely on having a particular representation such as the Burckhardt closed-form for the nonlinear friction coefficient function $\mu(\cdot)$. Indeed, whenever no knowledge of $\mu(\cdot)$ is available, the adversary can set $\hat{\mu}(\lambda)$ to be equal to zero in~\eqref{eq:dobDyn}.   This NDOB-based disturbance compensation technique is unlike the adaptive algorithms in~\cite{de2012torque,li2018hierarchical} where a particular representation of the friction coefficient function is needed and several parameters need to get updated simultaneously. As it will be seen in the simulations, even when $\hat{\mu}(\lambda)$ is set to zero, corresponding to a complete lack of knowledge by the adversary, the attack policy using the NDOB will meet its objectives. 
\end{remark}

     \section{Simulation Results}
\label{sec:sims}
In this section we present several numerical simulation results associated with \emph{five different attack policies} during braking in a straight road on both wet and dry asphalt. The simulation parameters are given in Table~\ref{tab:simTab}, where the quarter-car parameters are directly adopted from~\cite{de2012torque}.

Since the adversary would like to 
induce an almost complete wheel lockup condition during braking, it is desired that the trajectories $(v(t),\lambda(t))$ of the vehicle nonlinear traction dynamics in~\eqref{eq:tractionDyn2} converge to a very near vicinity of the lockup manifold $\mathcal{W}^b_L$ defined in~\eqref{eq:lockupManifold} within a relatively small amount of time (here, $T_c=0.95$ seconds).  
Out of the five attack policies employed by the adversary, one of them corresponds to applying a constant brake torque to the wheel, which is indeed a naive attack based on the assumption that with a relatively large brake torque the adversary can induce lockup in the wheels. The other four attacks employ the presented predefined-time controllers in the paper, where two of them that are given by~\eqref{eq:attPolicyNoDob}, with $p=0.15$ and $T_c=0.95$, do not possess any NDOB-based dynamic compensation mechanism. On the other hand, the last two predefined-time controllers, with $p=0.15$ and $T_c=0.95$, are employing the control policy in~\eqref{eq:attPolicyDob} with the disturbance estimate generated by the NDOB given by~\eqref{eq:attPolicyDob} with $L_d=2.65$.

 The nonlinear friction coefficient function is modeled using the three-parameter Burckhardt model (see, e.g.,~\cite{de1999model}) where 
\begin{equation}
\mu(\lambda) = c_1 (1-\exp(-c_2\lambda)) - c_3 \lambda.
\end{equation}
It is assumed that the adversary \emph{has no knowledge} of the nonlinear friction coefficient function. Accordingly, in all of the four non-constant adopted attack policies, $\hat{\mu}(\lambda)$ is set equal to zero. In Figure~\ref{tab:simTab}, the coefficients associated with dry and wet asphalt road conditions are given by $c_i$ and $c_i^\prime$, $1 \leq i \leq 3$, respectively. 
Furthermore, it is assumed that the adversary \emph{has no knowledge} of either the friction brake time constant $\tau_f$ or the friction brake deadtime $\delta_f$. Finally, it is assumed that the adversary \emph{has no knowledge} of the lower bounds $k^{\ast\prime}$ and $k^\ast$  in Propositions~\ref{prop:prop1} and~\ref{prop:prop2}. Therefore, $k$ in all four cases is set equal to zero. Finally, in the presented simulations, the external disturbances not related to road-tire interaction forces, i.e., $\Delta_v(t,v)$ and $\Delta_w(t,\omega)$, are set equal to zero. 

\begin{figure*}
	\hspace{-5ex}
	\begin{subfigure}{0.64\textwidth}
	\includegraphics[scale=0.23]{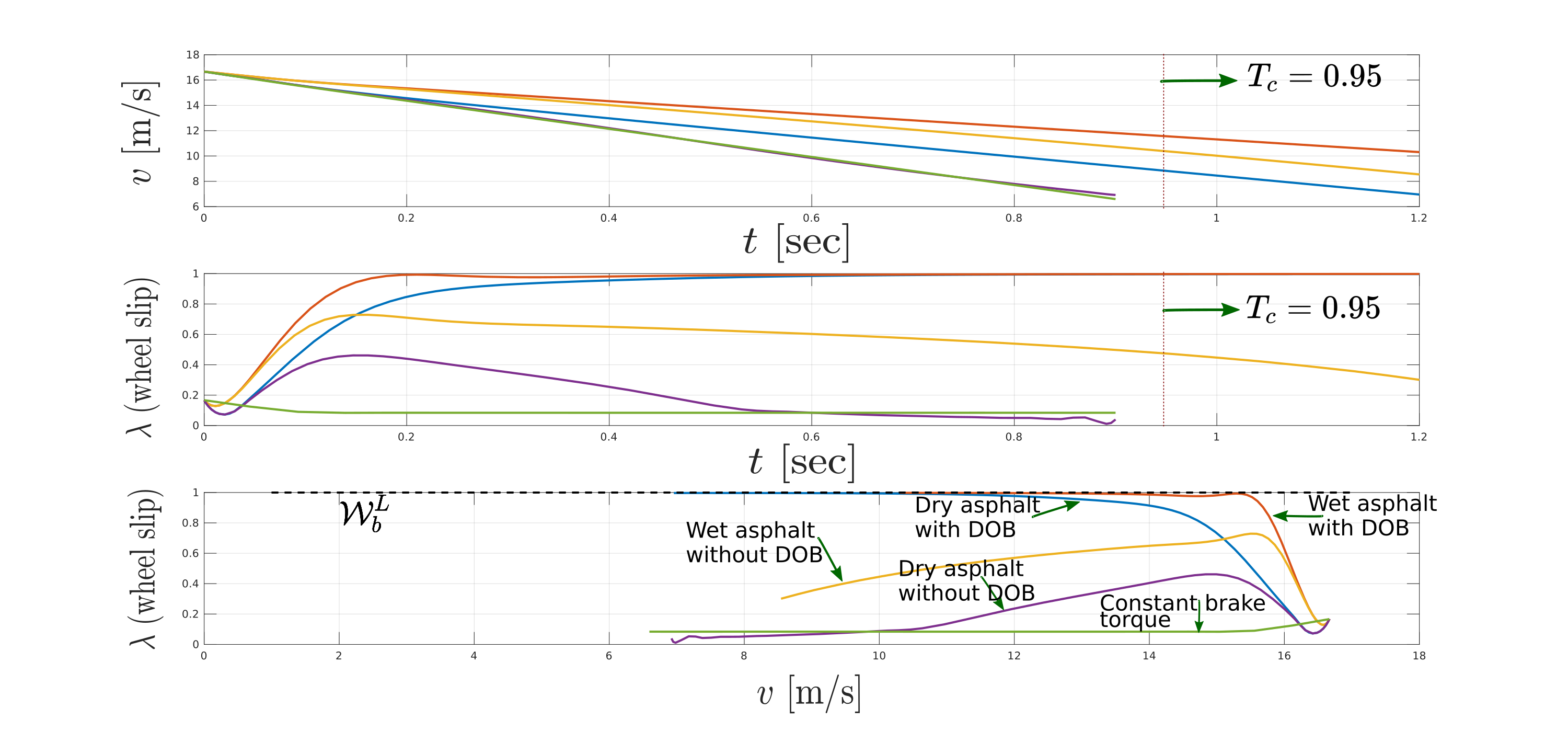} 
	\caption{}
	\end{subfigure}
%	\quad
    \hspace{-5ex}
	\begin{subfigure}{0.32\textwidth}
		\includegraphics[scale=0.16]{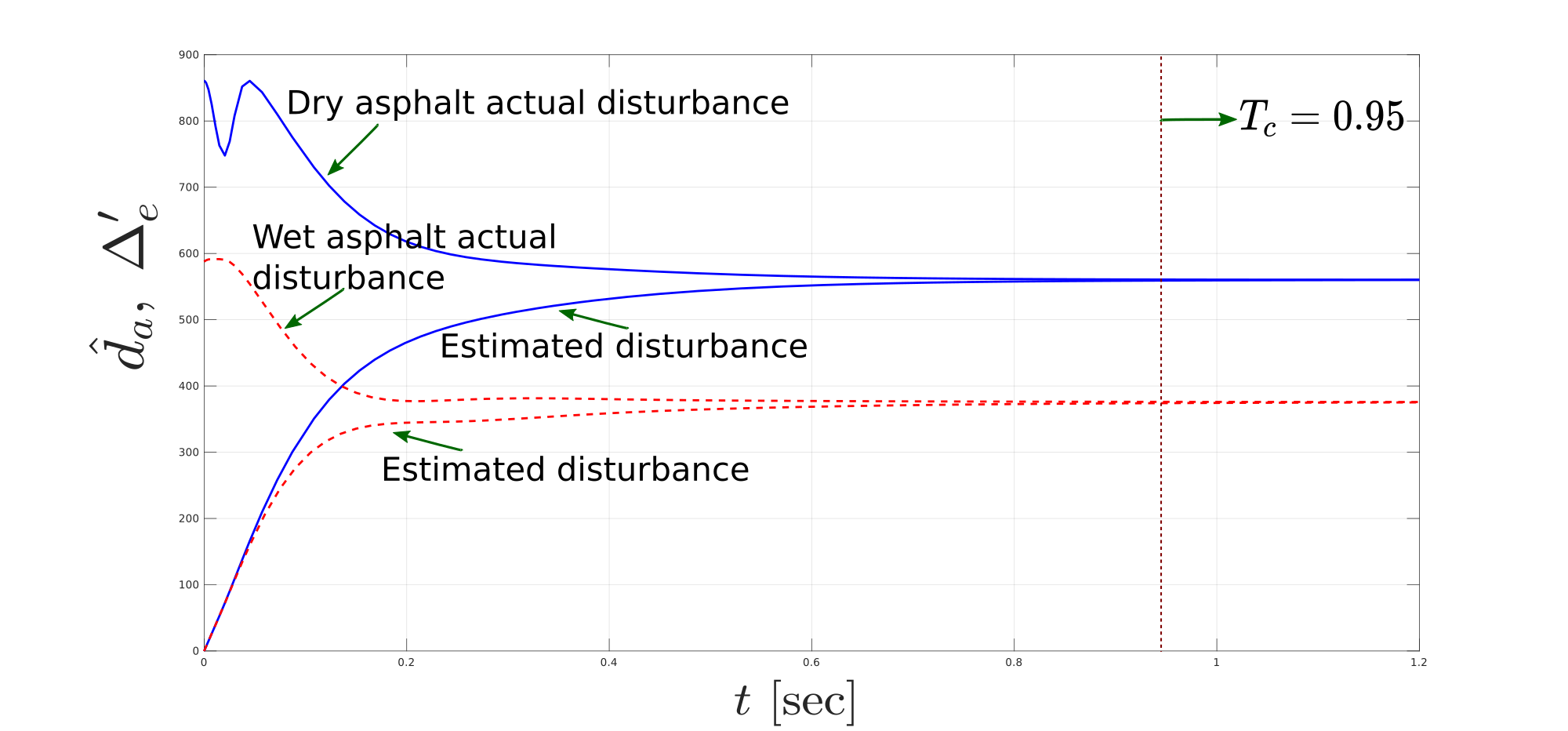} 
		\caption{}
	\end{subfigure}
	\caption{Profiles of the simulation results: (a) speed and wheel slip profiles under all five attack scenarios; (b) estimated disturbance $\hat{d}_a$ and the 
		actual disturbance $\Delta^\prime_e$ acting on the wheel slip error dynamics in the last two NDOB-based attack policies.}
	\vspace{-1ex}
	\label{fig:simTimeProf}
\end{figure*}

Figure~\ref{fig:simTimeProf} depicts the speed, wheel slip, and disturbance profiles from the simulations. As it can be seen from the Figure, in the three scenarios where the adversary does not employ the NDOB in~\eqref{eq:attPolicyNoDob3}, the attack objective is not met. We remark that had the adversary known an approximate representation of the tire-road interaction characteristics as well as the lower bounds $k^{\ast\prime}$ and $k^\ast$  in Propositions~\ref{prop:prop1} and~\ref{prop:prop2} (as opposed to setting all of them equal to zero), we still would have expected that the predefined-time attack policies without NDOBs to be successful on their own. The NDOB here is playing its well-known ``add-on'' role described in the DOB design literature~\cite{li2014disturbance,chen2004disturbance}. Indeed, it is \emph{the interplay in-between the predefined-time controller and the NDOB that makes the proposed braking attack policy less dependent on having a proper adversarial knowledge of the tire-road interaction characteristics.} 

%\begin{figure}
%	\centering
%	\begin{subfigure}{0.4\textwidth}
%%		\includegraphics[width=0.98\textwidth]{./figures/ODS_half_path.png} 
%		\caption{}
%	\end{subfigure}
%	\quad\quad
%	\begin{subfigure}{0.4\textwidth}
%	%	\includegraphics[width=0.98\textwidth]{./figures/ODS_half_u.png} 
%		\caption{}
%	\end{subfigure}
%	\begin{subfigure}{0.4\textwidth}
%%		\includegraphics[width=0.98\textwidth]{./figures/ODS_fourth_path.png} 
%		\caption{}
%	\end{subfigure}
%	\quad\quad
%	\begin{subfigure}{0.4\textwidth}
%	%	\includegraphics[width=0.98\textwidth]{./figures/ODS_fourth_u.png} 
%		\caption{}
%	\end{subfigure}
%%	\caption{Microrobot position and control input 
%%		time profiles under the ODS-based 
%%		scheme with: (a, b) $f_\text{SO}=12$ [Hz] and (c, d) 
%%		$f_\text{SO}=6$ [Hz].}
%	\label{fig:path_nominal2}
%\end{figure}
	
%	\begin{subfigure}{0.4\textwidth}
%		\includegraphics[width=0.98\textwidth]{./figures/fig2l.png} 
%		\caption{}
%	\end{subfigure}
%    \begin{subfigure}{0.4\textwidth}
%    	\includegraphics[width=0.98\textwidth]{./figures/fig2pd.png} 
%    	\caption{}
%    \end{subfigure}

%	\label{fig:path}
%\end{figure*}
%     
     \section{Concluding Remarks and Future Research Directions}
\label{sec:conc}
Motivated by the ample evidence in the automotive cybersecurity literature that the car brake ECUs can be maliciously reprogrammed,  
this paper demonstrated that an adversary with a limited knowledge of the tire-road interaction characteristics has the capability of inducing lockup conditions on the vehicle wheels in a finite time. The proposed attack policy for the frictional brakes  
relies on employing a predefined-time controller and a nonlinear disturbance observer, which compensates for the adversary's lack of knowledge of the tire-road interaction dynamics. This line of investigation on generating vehicle brake attack policies leads us to further research avenues. First,  this paper provides insights for the emerging area of attack generation against platoons of vehicles where string stability of a given platoon is of crucial importance. Second, this paper provides an urgent motivation for devising defensive ABS control policies that can protect the vehicle traction dynamics against such wheel lockup attacks. Finally, to have a better understanding of the safety implications under the proposed brake attack policies, the stability of vehicle lateral dynamics under such attacks needs to be thoroughly analyzed.

%   \input{./section/zerodyn}
%    \input{./section/CLFQP}
%\section*{ACKNOWLEDGMENTS}
%~~~~~~~~~~~~~~~~~~~~~~~~~~~~~~~~~~~~~~~~~	
\bibliographystyle{./styles/IEEEtran}
\bibliography{SaTCBib}

% Generated by IEEEtran.bst, version: 1.13 (2008/09/30)
\begin{thebibliography}{10}
\providecommand{\url}[1]{#1}
\csname url@samestyle\endcsname
\providecommand{\newblock}{\relax}
\providecommand{\bibinfo}[2]{#2}
\providecommand{\BIBentrySTDinterwordspacing}{\spaceskip=0pt\relax}
\providecommand{\BIBentryALTinterwordstretchfactor}{4}
\providecommand{\BIBentryALTinterwordspacing}{\spaceskip=\fontdimen2\font plus
\BIBentryALTinterwordstretchfactor\fontdimen3\font minus
  \fontdimen4\font\relax}
\providecommand{\BIBforeignlanguage}[2]{{%
\expandafter\ifx\csname l@#1\endcsname\relax
\typeout{** WARNING: IEEEtran.bst: No hyphenation pattern has been}%
\typeout{** loaded for the language `#1'. Using the pattern for}%
\typeout{** the default language instead.}%
\else
\language=\csname l@#1\endcsname
\fi
#2}}
\providecommand{\BIBdecl}{\relax}
\BIBdecl

\bibitem{huang2018vehicle}
J.~Huang, M.~Zhao, Y.~Zhou, and C.-C. Xing, ``In-vehicle networking: Protocols,
  challenges, and solutions,'' \emph{IEEE Network}, vol.~33, no.~1, pp. 92--98,
  2018.

\bibitem{kim2021cybersecurity}
K.~Kim, J.~S. Kim, S.~Jeong, J.-H. Park, and H.~K. Kim, ``Cybersecurity for
  autonomous vehicles: Review of attacks and defense,'' \emph{Comput. Secur.},
  p. 102150, 2021.

\bibitem{avatef2017}
O.~Avatefipour and H.~Malik, ``State-of-the-art survey on in-vehicle network
  communication can-bus security and vulnerabilities,'' \emph{Int. J. Comput.
  Sci. Netw.}, pp. 720--727, 2017.

\bibitem{checkoway2011comprehensive}
S.~Checkoway, D.~McCoy, B.~Kantor, D.~Anderson, H.~Shacham, S.~Savage,
  K.~Koscher, A.~Czeskis, F.~Roesner, T.~Kohno \emph{et~al.}, ``Comprehensive
  experimental analyses of automotive attack surfaces.'' in \emph{{USENIX}
  Secur. Symp.}, vol.~4.\hskip 1em plus 0.5em minus 0.4em\relax San Francisco,
  2011, pp. 447--462.

\bibitem{koscher2020experimental}
K.~Koscher, A.~Czeskis, F.~Roesner, S.~Patel, T.~Kohno, S.~Checkoway, D.~McCoy,
  B.~Kantor, D.~Anderson, H.~Shacham \emph{et~al.}, ``Experimental security
  analysis of a modern automobile,'' in \emph{The Ethics of Information
  Technologies}.\hskip 1em plus 0.5em minus 0.4em\relax Routledge, 2020, pp.
  119--134.

\bibitem{miller2015remote}
C.~Miller and C.~Valasek, ``Remote exploitation of an unaltered passenger
  vehicle,'' \emph{Black Hat USA}, vol. 2015, no. S 91, 2015.

\bibitem{froschle2017analyzing}
S.~Fr{\"o}schle and A.~St{\"u}hring, ``Analyzing the capabilities of the can
  attacker,'' in \emph{Eur. Symp. Res. Comput. Secur.}, 2017, pp. 464--482.

\bibitem{gillespie1992fundamentals}
T.~D. Gillespie, \emph{Fundamentals of vehicle dynamics}.\hskip 1em plus 0.5em
  minus 0.4em\relax Society of automotive engineers Warrendale, PA, 1992, vol.
  400.

\bibitem{giummarra2021classification}
M.~J. Giummarra, B.~Beck, and B.~J. Gabbe, ``Classification of road traffic
  injury collision characteristics using text mining analysis: Implications for
  road injury prevention,'' \emph{PloS one}, vol.~16, no.~1, p. e0245636, 2021.

\bibitem{teixeira2015secure}
A.~Teixeira, I.~Shames, H.~Sandberg, and K.~H. Johansson, ``A secure control
  framework for resource-limited adversaries,'' \emph{Automatica}, vol.~51, pp.
  135--148, 2015.

\bibitem{hoehn2016detection}
A.~Hoehn and P.~Zhang, ``Detection of covert attacks and zero dynamics attacks
  in cyber-physical systems,'' in \emph{2016 American Contr. Conf.
  (ACC)}.\hskip 1em plus 0.5em minus 0.4em\relax IEEE, 2016, pp. 302--307.

\bibitem{park2016adversary}
G.~Park, H.~Shim, C.~Lee, Y.~Eun, and K.~H. Johansson, ``When adversary
  encounters uncertain cyber-physical systems: Robust zero-dynamics attack with
  disclosure resources,'' in \emph{2016 IEEE 55th Conf. Dec. Contr. (CDC)},
  2016, pp. 5085--5090.

\bibitem{park2019stealthy}
G.~Park, C.~Lee, H.~Shim, Y.~Eun, and K.~H. Johansson, ``Stealthy adversaries
  against uncertain cyber-physical systems: Threat of robust zero-dynamics
  attack,'' \emph{IEEE Trans. Automat. Contr.}, vol.~64, no.~12, pp.
  4907--4919, 2019.

\bibitem{kontouras2017impact}
E.~Kontouras, A.~Tzes, and L.~Dritsas, ``Impact analysis of a bias injection
  cyber-attack on a power plant,'' \emph{IFAC-PapersOnLine}, vol.~50, no.~1,
  pp. 11\,094--11\,099, 2017.

\bibitem{wang2020detection}
X.~Wang, X.~Luo, X.~Pan, and X.~Guan, ``Detection and location of bias load
  injection attack in smart grid via robust adaptive observer,'' \emph{IEEE
  Syst. J.}, vol.~14, no.~3, pp. 4454--4465, 2020.

\bibitem{sanchez2015predefined}
J.~D. S{\'a}nchez-Torres, E.~N. Sanchez, and A.~G. Loukianov, ``Predefined-time
  stability of dynamical systems with sliding modes,'' in \emph{2015 American
  Contr. Conf. (ACC)}, 2015, pp. 5842--5846.

\bibitem{li2014disturbance}
S.~Li, J.~Yang, W.-H. Chen, and X.~Chen, \emph{Disturbance observer-based
  control: methods and applications}.\hskip 1em plus 0.5em minus 0.4em\relax
  CRC press, 2014.

\bibitem{chen2004disturbance}
W.-H. Chen, ``Disturbance observer based control for nonlinear systems,''
  \emph{IEEE/ASME Trans. Mechatron.}, vol.~9, no.~4, pp. 706--710, 2004.

\bibitem{mohammadi2017nonlinear}
A.~Mohammadi, H.~J. Marquez, and M.~Tavakoli, ``Nonlinear disturbance
  observers: Design and applications to {E}uler-{L}agrange systems,''
  \emph{IEEE Contr. Syst.}, vol.~37, no.~4, pp. 50--72, 2017.

\bibitem{mohammadi2018hybrid}
A.~Mohammadi, S.~Fakoorian, J.~C. Horn, D.~Simon, and R.~D. Gregg, ``Hybrid
  nonlinear disturbance observer design for underactuated bipedal robots,'' in
  \emph{2018 IEEE 58th Conf. Dec. Contr. (CDC)}, 2018, pp. 1217--1224.

\bibitem{rigatos2021flatness}
G.~Rigatos, N.~Zervos, P.~Siano, P.~Wira, and M.~Abbaszadeh, ``Flatness-based
  control for steam-turbine power generation units using a disturbance
  observer,'' \emph{IET Electr. Power Appl.}, doi: 10.1049/elp2.12077.

\bibitem{park2018stealthiness}
G.~Park, C.~Lee, and H.~Shim, ``On stealthiness of zero-dynamics attacks
  against uncertain nonlinear systems: A case study with quadruple-tank
  process,'' in \emph{Int. Symp. Math. Theory Netw. Syst. (ISMTNS)}, 2018, pp.
  10--17.

\bibitem{merco2020hybrid}
R.~Merco, F.~Ferrante, and P.~Pisu, ``A hybrid controller for {DOS}-resilient
  string-stable vehicle platoons,'' \emph{IEEE Trans. Intell. Transp. Syst.},
  vol.~22, no.~3, pp. 1697--1707, 2021.

\bibitem{polyakov2014stability}
A.~Polyakov and L.~Fridman, ``Stability notions and {L}yapunov functions for
  sliding mode control systems,'' \emph{J. Franklin Inst.}, vol. 351, no.~4,
  pp. 1831--1865, 2014.

\bibitem{zhou2016asymptotic}
B.~Zhou, ``On asymptotic stability of linear time-varying systems,''
  \emph{Automatica}, vol.~68, pp. 266--276, 2016.

\bibitem{polyakov2015finite}
A.~Polyakov, D.~Efimov, and W.~Perruquetti, ``Finite-time and fixed-time
  stabilization: Implicit {L}yapunov function approach,'' \emph{Automatica},
  vol.~51, pp. 332--340, 2015.

\bibitem{song2019time}
Y.~Song, Y.~Wang, and M.~Krstic, ``Time-varying feedback for stabilization in
  prescribed finite time,'' \emph{Int. J. Robust Nonlin. Contr.}, vol.~29,
  no.~3, pp. 618--633, 2019.

\bibitem{olson2003nonlinear}
B.~Olson, S.~Shaw, and G.~St{\'e}p{\'a}n, ``Nonlinear dynamics of vehicle
  traction,'' \emph{Veh. Syst. Dyn.}, vol.~40, no.~6, pp. 377--399, 2003.

\bibitem{johansen2003gain}
T.~A. Johansen, I.~Petersen, J.~Kalkkuhl, and J.~Ludemann, ``Gain-scheduled
  wheel slip control in automotive brake systems,'' \emph{IEEE Trans. Contr.
  Syst. Technol.}, vol.~11, no.~6, pp. 799--811, 2003.

\bibitem{de2012torque}
R.~De~Castro, R.~E. Ara{\'u}jo, M.~Tanelli, S.~M. Savaresi, and D.~Freitas,
  ``Torque blending and wheel slip control in {EV}s with in-wheel motors,''
  \emph{Veh. Syst. Dyn.}, vol.~50, no. sup1, pp. 71--94, 2012.

\bibitem{li2018hierarchical}
W.~Li, X.~Zhu, and J.~Ju, ``Hierarchical braking torque control of
  in-wheel-motor-driven electric vehicles over {CAN},'' \emph{IEEE Access},
  vol.~6, pp. 65\,189--65\,198, 2018.

\bibitem{de2011optimal}
R.~De~Castro, R.~Araujo, and D.~Freitas, ``Optimal linear parameterization for
  on-line estimation of tire-road friction,'' \emph{IFAC Proc. Vol.}, no.~1,
  pp. 8409--8414, 2011.

\bibitem{nie2017free}
S.~Nie, L.~Liu, and Y.~Du, ``Free-fall: Hacking {T}esla from wireless to {CAN}
  bus,'' \emph{Briefing, Black Hat USA}, vol.~25, pp. 1--16, 2017.

\bibitem{iso14229}
\BIBentryALTinterwordspacing
I.~O. for Standardization~(ISO). (2020) {ISO 14229-1:2020 Road vehicles —
  Unified diagnostic services (UDS) — Part 1: Application layer}. [Online].
  Available: \url{https://www.iso.org/standard/72439.html}
\BIBentrySTDinterwordspacing

\bibitem{de1999model}
C.~C. De~Wit, R.~Horowitz, and P.~Tsiotras, ``Model-based observers for
  tire/road contact friction prediction,'' in \emph{New Directions in nonlinear
  observer design}.\hskip 1em plus 0.5em minus 0.4em\relax Springer, 1999, pp.
  23--42.

\end{thebibliography}
%~~~~~~~~~~~~~~~~~~~~~~~~~~~~~~~~~~~~~~~~~	
\end{document}